\documentclass[twocolumn,nofootinbib]{revtex4-1}
\usepackage{graphicx}
\usepackage{epstopdf}
\usepackage{amsmath,amssymb}
\usepackage{url}

\DeclareMathOperator\im{\mathbb{I}\mathrm{m}}

\usepackage[shortlabels]{enumitem}
\begin{document}
\title{Photons from relativistic nuclear collisions}
\author{Hannah~Vormann$^{a}$}
\author{Tom~Reichert$^{a,b}$}
\author{Christian~Spieles$^{a}$}
\author{Jan~Steinheimer$^{e}$}
\author{Marcus~Bleicher$^{a,b,c,d}$}
\affiliation{$^a$ Institut f\"ur Theoretische Physik, 
Johann Wolfgang Goethe-Universit\"at, Max-von-Laue-Strasse 1, 60438 Frankfurt am Main, Germany}
\affiliation{$^b$ Helmholtz Research Academy Hesse for FAIR (HFHF), Campus Frankfurt, Max-von-Laue-Str. 12, 60438 Frankfurt am Main, Germany}
\affiliation{$^c$ GSI Helmholtz Center, Planckstr.~1, 64291 Darmstadt, Germany}
\affiliation{$^d$ John von Neumann-Institut f\"ur Computing, Forschungszentrum J\"ulich, 52425 J\"ulich, Germany}
\affiliation{$^e$ Frankfurt Institute for Advanced Studies, Ruth-Moufang-Str. 1, D-60438 Frankfurt am Main, Germany}


\begin{abstract}
Collisions of atomic nuclei at relativistic velocities allow to recreate the conditions encountered in neutron stars or in the early universe micro-seconds after the Big Bang. These reactions are performed in today's largest accelerator facilities, e.g. at CERN in Geneva, at the Relativistic Heavy Ion Collider at Brookhaven, NY or in the planned FAIR facility in Darmstadt Germany. During such a collision the matter is heated up to hundreds of MeV (billions of degrees) and compressed to densities of $3-10$ times the density inside ordinary atomic nuclei (i.e. $10^{17}-10^{18}$ kg/m$^3$). Usually these collisions are studied via the measurement of a multitude of strongly interacting particles, called hadrons, that are emitted at the end of the collision. However, also some photons are created. These photons are of special interest as they allow to look into the early stage of the collisions. This is because they are only interacting electro-magnetically with the hadrons of the created fireball which is a much weaker interaction compared to the strong interaction. This paper elucidates the physics of the photons and what can be learned from them.
\end{abstract}
\maketitle

%%%%%%%%%%%%%%%%%%%%%%%%%%%%%%%%%%%%%%%%%%%%%%%%

\section{Introduction}
One of the most fascinating achievements in the last 50 years was the development of accelerator technology to recreate the super hot matter present a microsecond after the Big Bang or the extreme densities that are only found in the interior of Neutron Stars. The largest of these particle accelerators is CERN's Large Hadron Collider (LHC), slightly smaller facilities exist in the US, the Relativistic Heavy Ion Collider (RHIC) or are planned, e.g. in Germany with the Facility for Anti-Proton and Ion Research (FAIR). These facilities accelerate atomic nuclei (often Gold or Lead nuclei) to velocities near the speed of light and smash them together. By this, one hopes to gain new insights into the structure of matter below the scale of protons and nucleons, namely into the dynamics of the Quarks and Gluons. The dynamics and properties of these Quarks and Gluons, the building blocks of protons and neutrons (and all other hadrons), are governed by the theory of strong interaction, also called Quantum Chromo-Dynamics (QCD).

Unfortunately, Quarks and Gluons can not be directly observed as they are bound to hadrons by a phenomenon called confinement. As the name ``strong interaction" already implies these hadrons interact intensely with each other and may even be seen as a form of liquid \cite{Muller:2007rs}. Indeed, the experiments at RHIC have hinted that such matter has the lowest viscosity ever observed. Unfortunately, this is a major disadvantage when one wants to obtain information from the interior of the collision zone: Because the strongly interacting hadrons are only emitted from the very late and very dilute stage of the system's evolution, when the interactions cease. Thus, they carry mainly information from the end of the reaction when the system has already cooled down. 

How can one then obtain a deeper view into the most interesting hot and dense region of the fireball created in such collisions? Here Photons, the quanta of light, are perfect candidates. The main reason is that photons do not interact strongly with the surrounding matter, but interact only via the electromagnetic interactions, which is typically a factor $\alpha=1/137$ smaller than the strong interaction. Thus, once a photon is created it can leave the fireball unscathed and can tell about the temperatures of the fireball when it is hottest and most dense. 

In this paper, we want to give some introduction into this exciting topic and want to present the steps needed for such calculations in a pedagogical manner. The paper is structured in the following way: First, we describe the hydrodynamic set-up used for the calculations. Here, we used a standard (3+1) dimensional relativistic hydrodynamic calculation, supplemented with a hadron resonance gas equation of state to close the equations. In the next section, we provide a short review on the motivation to use photons to explore the properties of QCD-matter. Here, we focus on the production of real photons, i.e. photons with vanishing mass (these are the photons one knows from daily experience, albeit the energy of the photons is higher than usual) and omit the discussion of virtual photons, i.e. massive photons which are also used to explore hot and dense QCD-matter (we refer the interested reader to \cite{vanHees:2007th,Rapp:2009yu}). Then we give a brief overview on how to calculate photon emission rates from hadronic collisions. For simplicity, we will only discuss the most important channel ($\pi\rho\rightarrow \pi\gamma$) and provide references for further detailed information. Finally, we show our results for the photon rates for three different energies. The paper concludes with a summary and a critical assessment of the omitted effects.

The whole study was carried out as part of a summer study program at the Frankfurt University over a duration of 8 weeks.

\section{Technical set-up for the hydrodynamic evolution and method of solution}
When one wants to explore the dynamics of QCD-matter one is confronted with a multitude of complications. QCD, being a non-abelian gauge theory, practically prohibits ab-initio calculation of dynamical processes due to the complexity of the theory. However, some direct information can be obtained from large scale lattice QCD simulations for static systems at very low baryon densities \cite{Philipsen:2007rj}. Here QCD predicts that the nuclear matter should undergo a transition from hadron degrees-of-freedom (protons, neutrons, pions, etc.) to quark and gluon degrees-of-freedom. The temperature of this transition depends on the baryon density and is around $T_c\approx 150-160$~MeV for vanishing baryon density. Thus, below the transition temperature the system consists of hadrons which interact and form the dense matter observed at moderate collision energies.

How does one then model the dynamics? Typically there are two choices: 
\begin{itemize}
    \item 
    A Boltzmann equation type simulation \cite{Bleicher:1999xi}, where one uses individual matrix elements to describe the interaction cross sections between the various hadron species or 
    \item
    a hydrodynamic calculation, where one assumes that the QCD matter can be treated as a liquid \cite{Kolb:2003dz}.
\end{itemize}

Both approaches have advantages and disadvantages: The Boltzmann equation can usually only be used for dilute systems and requires knowledge of all individual matrix elements, in contrast hydrodynamics is only applicable if the system is rather dense and can be assumed to be in local equilibrium. Instead of individual matrix elements, the properties of the fluid enter via an equation of state, i.e. the relation between pressure $p$ and energy density $\varepsilon$. Here, we choose relativistic hydrodynamics, because the calculational set-up and the numerical implementation is easier for the present task. 

Let us now review shortly the main ingredients into the hydrodynamic simulation. For a more detailed description, we refer the reader to the very pedagogical introduction to hydrodynamics of heavy ion collisions by Ollitrault \cite{Ollitrault:2007du} and references therein.

In general, hydrodynamics represents local energy-, local momentum- and local charge-conservation. The energy-momentum tensor $T^{\mu\nu}$ for an ideal fluid\footnote{Using an ideal fluid instead of a viscous fluid is a good approximation, because the viscosity of the matter produced in heavy ion collisions is sufficiently small (see \cite{Muller:2007rs,Reichert:2020oes} and references therein).} is 
\begin{equation}\label{Tmunu}
    T^{\mu\nu}=(\varepsilon+p)u^\mu u^\nu-g^{\mu\nu} p\, .
\end{equation}
Here and in the following, we use natural units, i.e. $\hbar=c=k_B=1$, Einstein sum convention is implicit and the west coast metric, (i.e. the spatial diagonal elements of $g^{\mu\nu}$ carry the minus sign), and $u^\mu$ is the velocity of the fluid cell while $p$ and $\varepsilon$ denote the cell's pressure and energy density, respectively.

The conservation of energy and momentum is described by
\begin{equation}\label{Tmunu_cons}
    \partial_\mu^{}T^{\mu\nu}_{}=0\, .
\end{equation}

In addition to the energy momentum conservation, we also include the conservation of the baryon charge, which is described via the conservation of the baryon current $j_B^\mu=\rho_B u^\mu$, with $\rho_B$ being the baryon density in the local rest frame of a hydrodynamic cell \cite{Ollitrault:2007du,Rischke:1998fq}.

The conservation equation for the baryon current is then
\begin{equation}\label{Nmu}
    \partial_\mu^{}j_B^\mu=0\, .
\end{equation}

To close the system of equations, we also need to specify the relation between energy density and pressure, $p(\varepsilon,\rho_B)$ or $p(T,\mu_B)$, where $T$ is the local temperature in the cell and $\mu_B$ is the chemical potential of the baryons in the cell needed to fix the baryon density. This relation is known as the equation of state (EoS) and defines the properties of the fluid. In \cite{Ollitrault:2007du} the simplifying assumption of a gas of massless particles was made, leading to a simple EoS of $p=1/3 \varepsilon$. For the low energies discussed here, one can not make this assumption, but has to employ a hadron resonance gas with massive particles. 

It was shown in \cite{Dashen:1969ep} that an interacting gas of massive hadrons can be approximated by treating unstable particles as explicit degrees of freedom in the partition function of the gas. This is known as the hadron resonance gas approximation.
In this approximation, the logarithm of the partition function ${\cal Z}$ is given as
\begin{eqnarray}
\ln {\cal Z} & = & \sum_i \ln {\cal Z}_i(T,V,m_i,\mu_i)
\nonumber \\ 
& = &
\sum_i (\pm 1) \frac{g_i V}{2\pi^2} 
\\
\nonumber && \qquad \times
 \int_0^\infty \mathrm{d}k\, k^2\, \ln \left ( 1 \pm e^{\mu_i/T} e^{-\sqrt{m_i+k^2_{}}/T} \right )\, ,
\end{eqnarray}
where the sum runs over all hadronic species in the particle data book \cite{ParticleDataGroup:2018ovx}. The upper (lower) signs are for fermionic (bosonic) species, $g_i$ is the spin and iso-spin degeneracy factor of the hadron, $m_i$ is the mass of the hadron, $V$ is the volume, and $\mu_i$ is the chemical potential of species $i$.

The pressure is then given by
\begin{equation}
    p(T,\mu_B)= \left.\frac{\partial\, T \ln {\cal Z} (T, V, \mu_B)}{\partial V}\right|_{T,\mu}\, .
\end{equation}

This can be done straightforward numerically and one obtains $p(T,\mu_B)$ for the hadron resonance gas in tabulated form. More elaborate Equations of State are available, see e.g. \cite{Motornenko:2019arp}. 

The hydrodynamic equations can be solved analytically for a system without baryon current and for a (1+1) dimensional geometry (one longitudinal direction plus time). This simple solution, known as the Bjorken solution \cite{Bjorken:1982qr} (see \cite{Ollitrault:2007du} for a more pedagogical presentation), is unfortunately only valid for very high collision energies. Towards lower energies, an approximate analytical solution for a (2+1) dimensional system (transverse and longitudinal direction plus time) was found by Belenkij and Landau \cite{Belenkij:1955pgn}. For the (3+1) dimensional case analytical solutions do not exist and one needs to solve the equations numerically.

For the present calculation we employ a standard numerical scheme to solve the equations on an Eulerian grid, known as the SHASTA (Sharp And Smooth Transport Algorithm) \cite{Boris:1973tjt,Rischke:1995ir}. SHASTA is an Eulerian finite-difference algorithm for solving the continuity equations employing a flux correction technique. The choice of the algorithm is a matter of taste and other solvers, e.g. of Gudanov type \cite{Karpenko:2013wva} or the Kurganov-Tadmor method \cite{KURGANOV2000241,Schenke:2010nt}, can also be applied. E.g. a hydrodynamical code package can be downloaded from \cite{Code}.

\subsection{Evolution of the matter}
%
%%%%%%%%%%%%%%%%%%%%%%%%%%%
\begin{figure}[t!]
\begin{center}
	\includegraphics[width=0.45\textwidth]{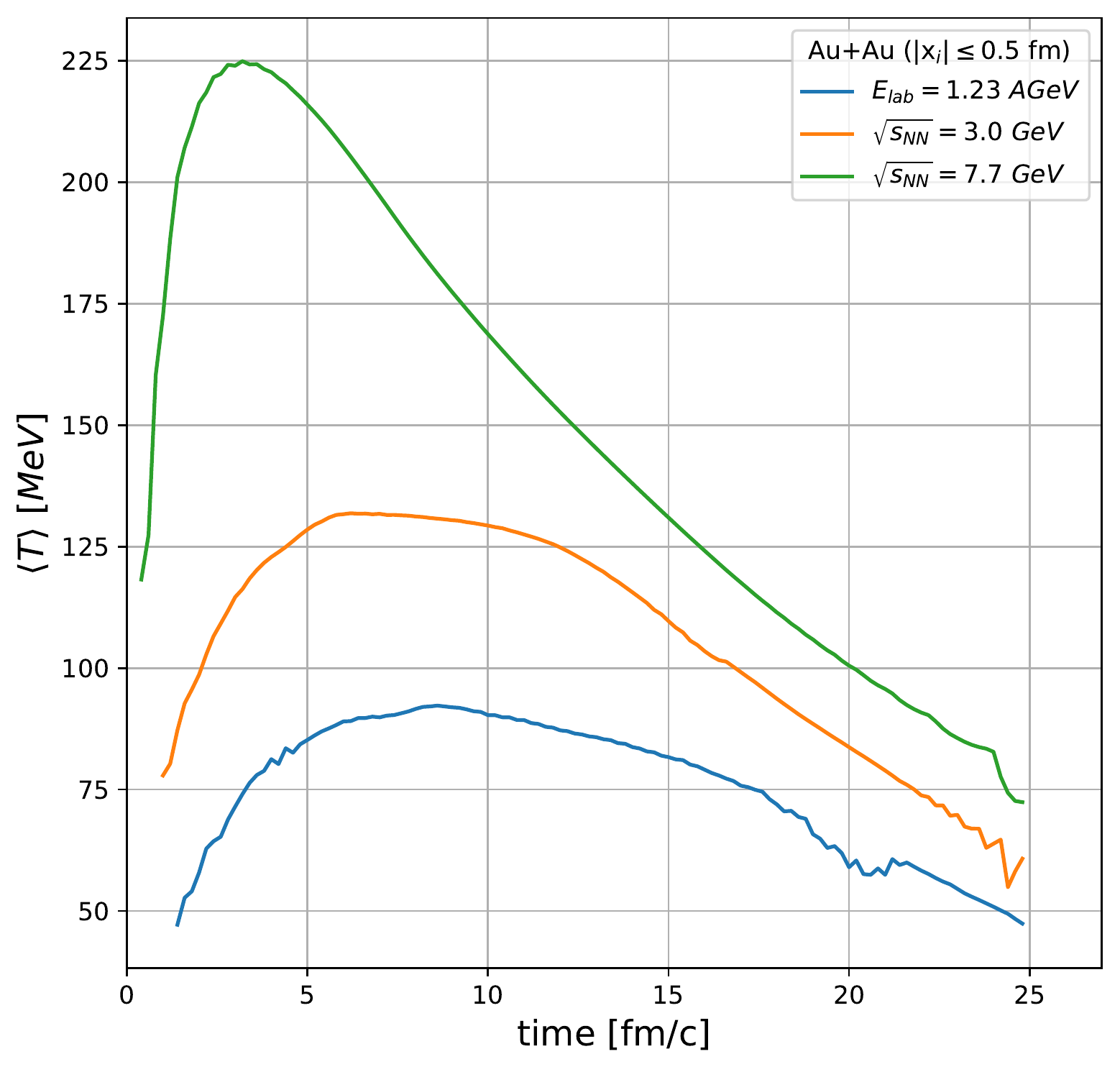}
\end{center}
\caption{[Color online] Evolution of the temperatures in the central region of the reaction for three selected beam energies in the GSI, RHIC-BES region of $E_{lab}=1.23 A$~GeV (blue), $\sqrt{s_{NN}}=3$~GeV (orange) and $7.7$~GeV (green).}
\label{fig:temp_evol} 
\end{figure}
%%%%%%%%%%%%%%%%%%%%%%%%

After the description of the methodological set-up, let us come to the physical set-up and physical assumptions used for the present investigation. As initial condition for the (3+1) dimensional simulation, we set two gold nuclei (target and projectile) on the hydrodynamic grid that approach each other in longitudinal direction. Each gold nucleus is modeled as a sphere of radius $r=1.2 A^{1/3}$, where $A$ is the mass number, and volume $V=4/3 \pi r^3$ and constant $\rho_B=A/V= 0.14$ fm$^3$. When the gold fluids from target and projectile come into contact they are immediately thermalized and their kinetic energy is translated into thermal energy. Here a word of caution is necessary: The assumption of instant local thermalization is rather crude and is only justified as long as the thermal motion (average thermal velocity) of the hadrons after the thermalization of the projectile and target fluids is comparable to the longitudinal velocities of the incoming nuclei. While this assumption is fulfilled at low energies, it becomes questionable at energies above $\sqrt{s_{NN}}\approx 5-10$~GeV. Here, one would need to employ multi-fluid simulations, which are however out of the scope of this investigation.

Let us now investigate the evolution of the temperatures encountered in gold+gold (from now on denoted as Au+Au) reactions at various energies currently studied experimentally at the Relativistic Heavy Ion collider (RHIC, New York) and at the SIS100 at Darmstadt, Germany. 
In Fig.~\ref{fig:temp_evol} we show the evolution of the temperatures in the central region of the reaction for three selected beam energies in the GSI, RHIC-BES region of $E_{lab}=1.23 A$~GeV, $\sqrt{s_{NN}}=3$~GeV and $7.7$~GeV. The central region here was  defined as a cube with a side length of 1~fm centered around the origin of the grid. We observe that the system heats up tremendously leading to peak temperatures in the order of $T_{max}=90$ MeV ($E_{lab}=1.23 A$~GeV), $T_{max}=130$~MeV ($\sqrt{s_{NN}}=3$~GeV) and even $T_{max}=225$~MeV. The maximum of the curves corresponds to the full overlap time of the nuclei, this is why it shifts in time to lower values when the collision energy increases (the nuclei become Lorentz contracted!).  

\begin{figure}[t!]
\begin{center}
	\includegraphics[width=0.45\textwidth]{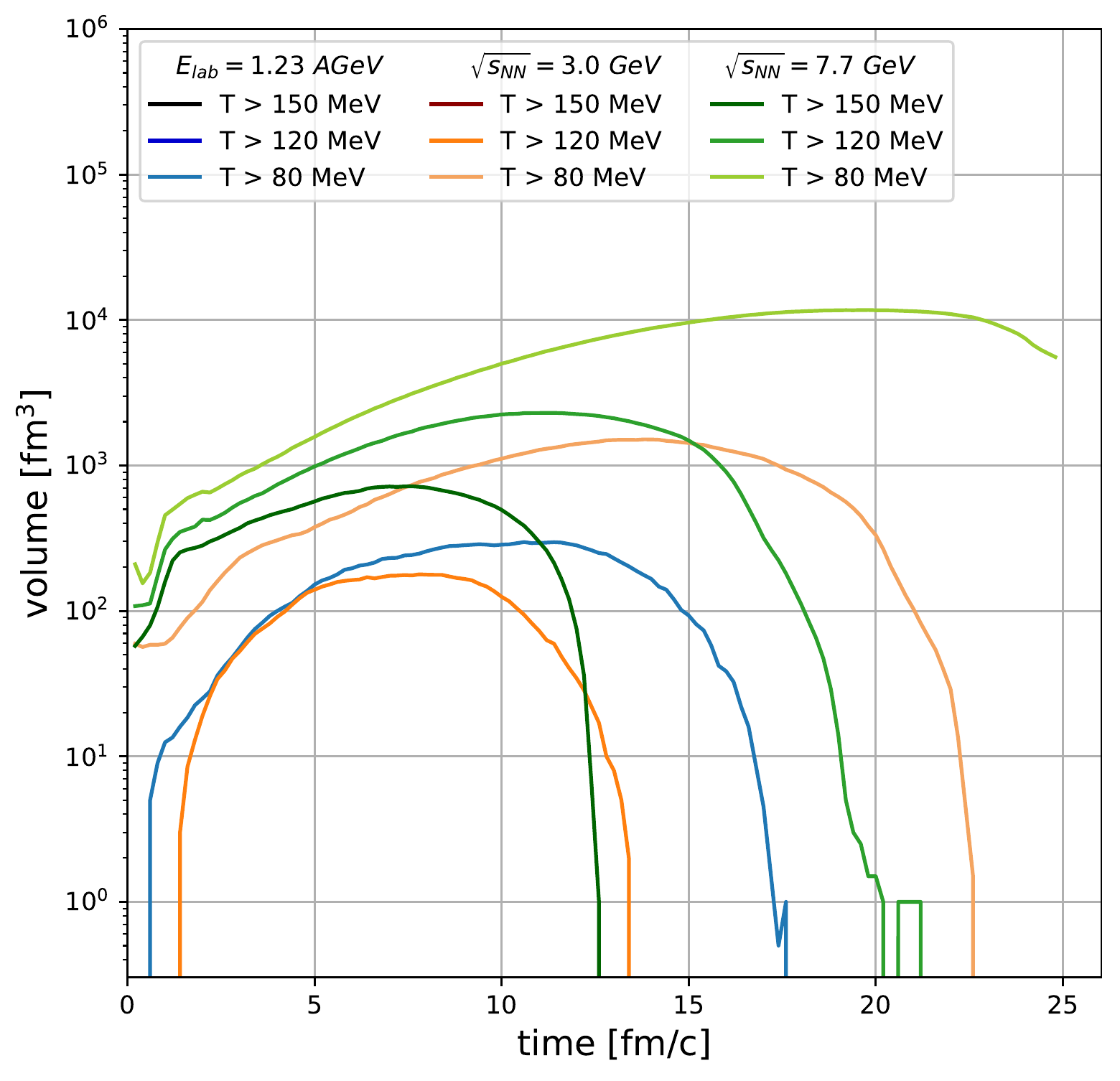}
\end{center}
\caption{[Color online] Time dependence of the volume of 'hot' cells, i.e. cells with a temperature above $T=80$~MeV, $T=120$~MeV, and $T=150$~MeV for three selected beam energies in the GSI, RHIC-BES region of $E_{lab}=1.23 A$~GeV (blue curves), $\sqrt{s_{NN}}=3$~GeV (orange curves) and $7.7$~GeV (green curves).}
\label{fig:volume_evol} 
\end{figure}

In Fig.~\ref{fig:volume_evol} we show the time dependence of the volume of ``hot" cells, i.e. cells with a temperature above $T=80$~MeV, $T=120$~MeV, and $T=150$~MeV for three selected beam energies in the GSI, RHIC-BES region of $E_{lab}=1.23 A$~GeV, $\sqrt{s_{NN}}=3$~GeV and $7.7$~GeV. We observe that the hottest regions of the system sustain only a short period of time during the evolution which rapidly cease. At the largest investigated beam energy the region with $T>80$~MeV becomes more long-lived and the corresponding volume is increasing until 20~fm/c. In the next section we will turn to direct photon production to see how the four-volume evolution of the temperature affects their production rates.

%%%%%%%%%%%%%%%%%%%%%%%%%%%
%%%%%%%%%%%%%%%%%%%%%%%%

\section{Photon production rates}
The collision of two nuclei leads to the formation of a hot and dense fireball due to the compression of nuclear matter and energy into a small volume in space. The lifetime of this fireball is defined as the time until a final state of freely streaming particles is reached as a result of the collective expansion of matter. The trajectory of the fireball in the T-$\mu_B$-plane of the QCD phase diagram depends on the energy of the collision. Above a critical temperature $T_c$ a Quark-Gluon-Plasma (QGP), a deconfined state of quarks and gluons, is expected to be formed. In order to probe possible phase transitions one needs suitable observables which do not only carry information about the final state after the freeze-out of the system but the whole evolution of the fireball especially in the initial hot and dense phase. Hadrons are  unsuited due to typically short mean free paths. This is because they interact strongly with the surrounding matter so that they only escape from the fireball's surface or after the expansion of the fireball when the scattering rate is much smaller than the expansion rate. Contrary, electromagnetic probes like photons and dileptons (virtual photons) are promising candidates: Their mean free paths are typically larger than the fireball itself meaning that they suffer little final state interaction (they do not underlay the strong interaction) and their emission rates rapidly vary with local intensive variables. In this paper we restrict ourselves to the emission of real photons.

The production mechanisms of photons can be divided into three types \cite{Arleo:2004gn}: Firstly, photons can be produced primordially in so called hard QCD interactions of two colliding nuclei (QCD Compton, annihilation, and bremsstrahlung). They are called ``prompt'' photons. Rates of these processes can be computed within the framework of perturbative QCD (pQCD) and dominate at large transverse momenta $p_T$. Secondly, when a QGP forms, photons are emitted from collisions of quarks and gluons. This radiation also dominates in the large $p_T$ regime. Thirdly, the QGP expands and reenters the hadronic phase. Here, hot hadronic resonances emit photons until the freeze-out of the system whose rates dominate at small $p_T$. Together with the QGP photons they are called ``thermal'' photons. Additionally, there are also ``decay'' photons which are decay products of hadronic resonances. Unbiased experimental spectra consist of the sum of all of these photons. In order to identify the ''thermal'' photons, other competing components have to be subtracted by the experiment. In the present paper we limit ourselves to real, thermal photons, that is photons from secondary interactions of particles in the expanding fireball, and focus on hadron gas emission.
For the calculation of photon emission rates we follow Turbide, Rapp and Gale (TRG) \cite{Turbide:2003si}. They present two different ways of defining thermal photon rates: In the context of thermal field theory the differential photon emission rate from an equilibrated system reads
\begin{align}
    \frac{\mathrm{d}R_\gamma}{\mathrm{d}^4p} = P_{\mu\nu} W^{\mu\nu}
\end{align}
with the polarization-summed photon tensor $P_{\mu\nu}$ and the hadronic tensor $W^{\mu\nu}(p;\mu_B,T)$ describing the effects of strong interaction. This can also be written as follows
\begin{align}
    p_0 \frac{\mathrm{d}R_\gamma}{\mathrm{d}^3p} = - \frac{\alpha}{\pi^2}\ f^B(p_0;T)\  \im\Pi^{\mu\nu}_{em}(p_0=p;T)
\end{align}
where $f^B$ is the Jüttner distribution and $\Pi^{\mu\nu}_{em}$ denotes the in-medium photon self-energy whose imaginary part is directly related to $W^{\mu\nu}$.

Alternatively, within relativistic kinetic theory the photon rate of a given reaction $1+2 \rightarrow 3+\gamma$ can be written as
\begin{align}
    \label{photon_rate_kinetic}
    p_0 \frac{\mathrm{d}R_\gamma}{\mathrm{d}^3p} = \int &\frac{\mathrm{d}^3p_1}{2(2\pi)^3E_1} \frac{\mathrm{d}^3p_2}{2(2\pi)^3E_2} \frac{\mathrm{d}^3p_3}{2(2\pi)^3E_3} \nonumber\\
    &\times\left( (2\pi)^4 \delta^{(4)}(p_1+p_2 \rightarrow p_3+p) \right. \nonumber\\
    &\times\left. |\mathcal{M}|^2 \frac{f(E_1) f(E_2) [1 \pm f(E_3)]}{2(2\pi)^3} \right)
\end{align}
with $\mathcal{M}$ being the invariant scattering matrix element for the reaction under consideration.

In the equations above, the emission rate is the rate of production per unit time and volume
\begin{align}
    \mathrm{d}R_\gamma = \frac{\mathrm{d}N_\gamma}{\mathrm{d}^4x}\ .
\end{align}
A meson gas consisting of pseudo-scalar, vector, and axial mesons ($\pi$, $K$, $\rho$, $K^*$, $a_1$) can be characterized using an effective massive Yang-Mills Lagrangian. For all interactions occurring in this Lagrangian that produce photons, the emission rate can be calculated using Eq.~\ref{photon_rate_kinetic} employing the respective matrix elements $\mathcal{M}$.

Additionally, so-called hadronic vertex form factors have to be applied to these rates to consider finite-size effects. This results in a suppression compared to the bare rates. According to TRG they can be written in the form
\begin{align}
    F(t) = \left( \frac{2 \Lambda^2}{2 \Lambda^2-t} \right)^2
\end{align}
where $\Lambda=1$~GeV denotes a cutoff parameter and $t$ is the four-momentum transfer in a given t-channel exchange.
Its average $\overline{t}$ can be approximated as
\begin{align}
    \overline{t} \simeq - 2 E m_\pi
\end{align}
for a t-channel pion exchange \cite{Arleo:2004gn}. Here, $E$ is the energy of the emitted photon and $m_\pi$ the pion mass.

With the help of this form factor TRG have parametrized the photon emission rates for different production channels in the non-strange and strange sector \cite{Turbide:2003si}. These parametrizations are used here in order to obtain thermal photon spectra of collisions of two gold atoms.

For the present calculation we use this parametrization for the $\pi+\rho\rightarrow \pi+\gamma$ reaction rate \cite{Turbide:2003si}:
\begin{widetext}
\begin{equation}
    E\frac{\mathrm{d}R_{\pi + \rho \rightarrow \pi+\gamma}}{\mathrm{d}^{3}p} = F^{4}(E)\, T^{2.8}\exp\left(\frac{-(1.461 T^{2.3094}+0.727)}{(2 T E)^{0.86}}+(0.566 T^{1.4094}-0.9957)\frac{E}{T}\right)\ [\mbox{fm}^{-4}\mbox{GeV}^{-2}]
    \label{eq:rate}
\end{equation}
\end{widetext}

In the non-strange sector the reaction $\pi \rho \rightarrow \pi \gamma$ turned out to be the dominating source of photons, in the strange sector it is $K^* \pi \rightarrow K \gamma$ with a contribution of approximately 20\% of the former. We neglect the less important channels to keep the study simple.

When comparing their data for hadron gas emissions to QCD calculations of QGP emissions, TRG could show a strong similarity of the two rates near the expected phase transition. This is known as parton-hadron duality and means that near the phase transition temperature both, quark and gluon degrees-of-freedom and hadron degrees-of-freedom yield similar results. We hence focus on the hadronic photon production in this paper and assume its validity also at higher temperatures.

\begin{figure}[t]
\begin{center}
	\includegraphics[width=0.45\textwidth]{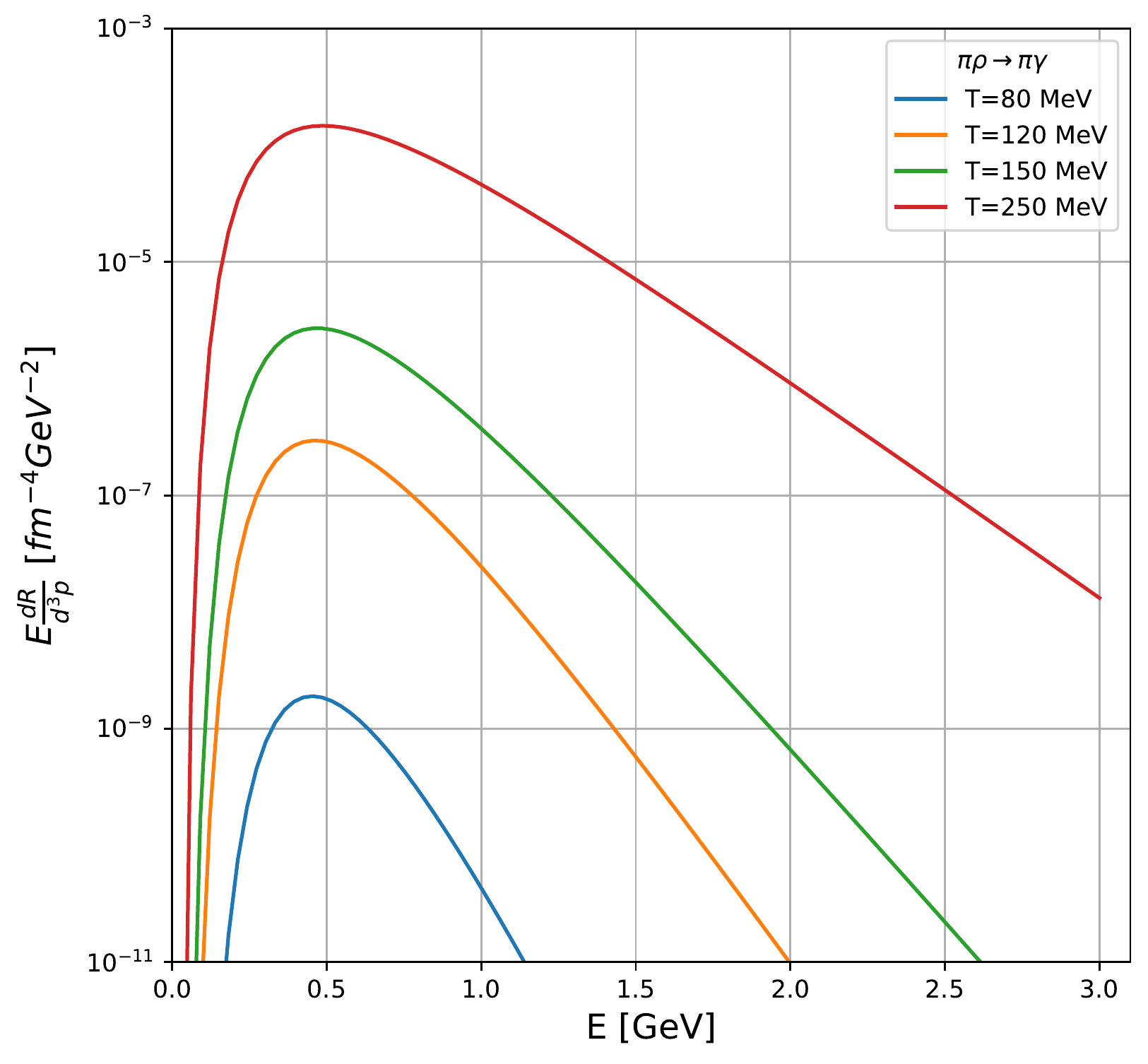}
\end{center}
\caption{[Color online] The rate of photon emission rate from the process $\pi+\rho\rightarrow \pi+\gamma$ from a thermal hadron gas at four different temperatures from 80 MeV to 250 MeV (from bottom to top).}
\label{fig:photon_rate} 
\end{figure}

In Fig.~\ref{fig:photon_rate} we show the rate of photon emission rate from the process $\pi+\rho\rightarrow \pi+\gamma$ from a thermal hadron gas at four different temperatures.

%%%%%%%%%%%%%%%%%%%%%%%%%%%
%%%%%%%%%%%%%%%%%%%%%%%%

\section{Results and discussion}
Now all ingredients are ready and we are able to calculate the photon emission from the simulation of a real collision of two heavy ions. The momentum differential photon yield (rate $R_i\times V^{cell}_i\times \Delta t$) is summed up for each individual cell $i$ on the hydrodynamic grid over the course of the reaction. Thus it integrates the photon emission over the whole volume and time of the fireball with different emission temperatures at each space-time point.

\begin{figure}[t!]
\begin{center}
	\includegraphics[width=0.45\textwidth]{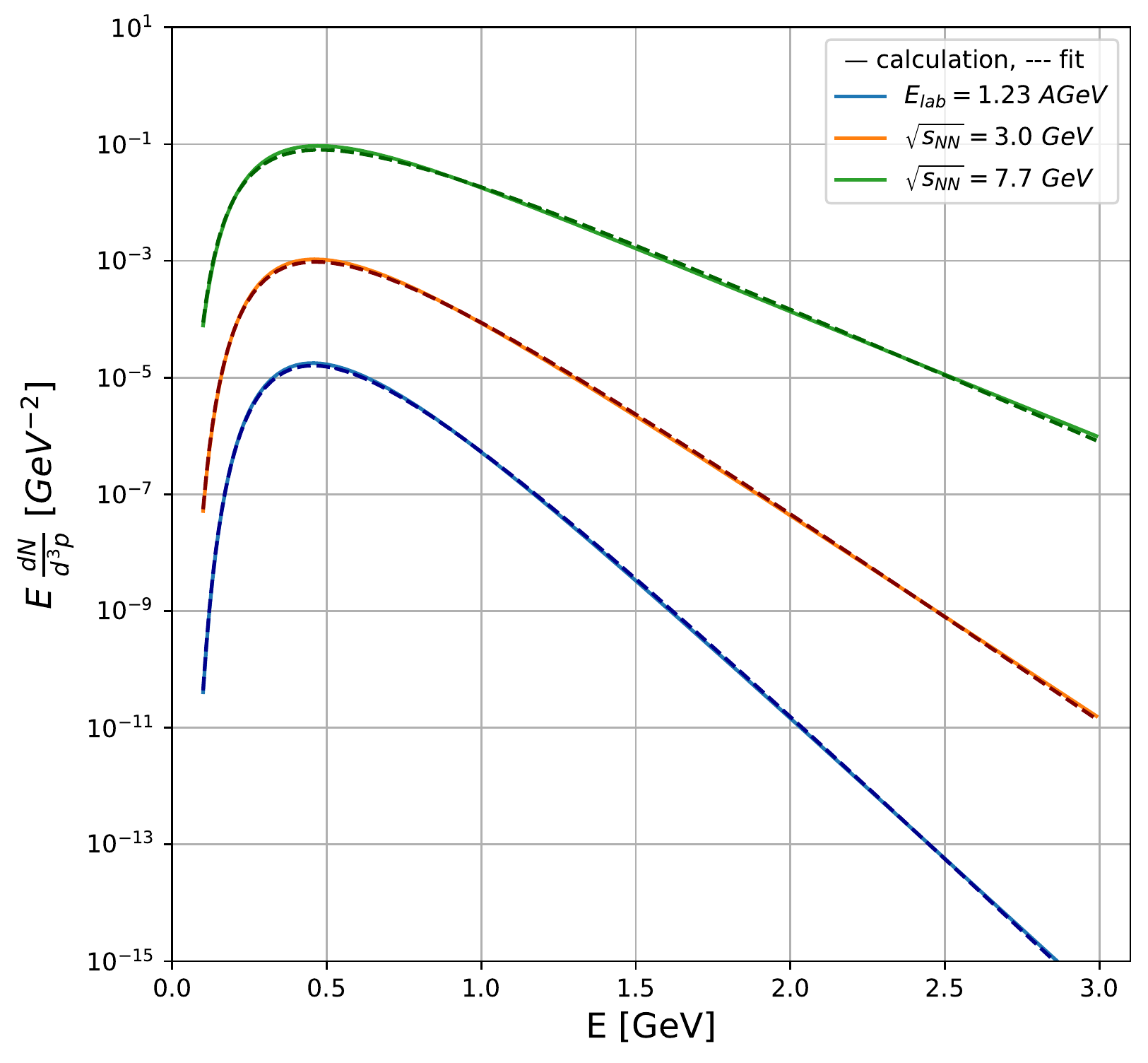}
\end{center}
\caption{[Color online] Invariant photon spectra for three selected beam energies in the GSI, RHIC-BES region of $E_{lab}=1.23 A$~GeV (blue), $\sqrt{ s_{NN}}=3$~GeV (orange) and $7.7$~GeV (green). The calculations are shown as full lines, the fits are shown as dotted lines.}
\label{fig:photon_spectra} 
\end{figure}

In Fig.~\ref{fig:photon_spectra} we show the invariant photon spectra for three selected beam energies in the GSI, RHIC-BES region of $E_{lab}=1.23 A$~GeV, $\sqrt{s_{NN}}=3$~GeV and $7.7$~GeV.
The first observation is that the number of photons increases strongly with increasing beam energy. The main reason for this behavior is the raise in temperature with the collision energy. This is in line with the rates in Fig.~\ref{fig:photon_rate} at fixed temperatures where an increase in temperature of about $30-40$~MeV leads to emission rates that are up to two orders of magnitude higher. Additionally, the space-time volume (volume $\times$ lifetime of the system) of the source increases with increasing energy. This is expected because at higher collision energies, more hadrons are produced and the fireball becomes larger and lives longer until it decays (cf. with Fig.~\ref{fig:volume_evol}). The second observation concerns the slope of the spectra. Here one observes that the spectra become ``hotter'' with increasing beam energy. That is also in line with our expectations that an increase in collision energy leads to a hotter temperature.

Thus, the photon spectrum allows us to extract two important informations about the collision: 1) the space time volume of the hot region and 2) the average temperature during the course of the reaction. To show this, we have fitted the curves in  Fig.~\ref{fig:photon_spectra} (shown as dotted lines) with the photon yield, Eq.~\ref{eq:rate}, and extracted the 4-volume of the emission source and the average temperature. At $E_{lab}=1.23 A$~GeV we obtain $V_4=$ volume $\times$ lifetime of the system $=2014$~fm$^4$ and $T=89$~MeV, at $\sqrt {s_{NN}}=3$~GeV we obtain $V_4=2197$~fm$^4$ and $T=125$~MeV and at $\sqrt {s_{NN}}=7.7$~GeV we obtain $V_4=2652$~fm$^4$ and $T=200$~MeV. The obtained 4-volumes are consistent with the volume evolution provided in Fig.~\ref{fig:volume_evol} using these averaged temperatures. It is also interesting to observe that the averaged temperatures cross the QCD deconfinement temperature in the energy region around $\sqrt{s_{NN}}=3-7.7$ GeV. This shows the power of the analysis of photon spectra in contrast to hadron spectra as they allow to observe the hottest regions of the fireball.

%%%%%%%%%%%%%%%%%%%%%%%%%%%
%%%%%%%%%%%%%%%%%%%%%%%%

\begin{figure}[t!]
\begin{center}
	\includegraphics[width=0.45\textwidth]{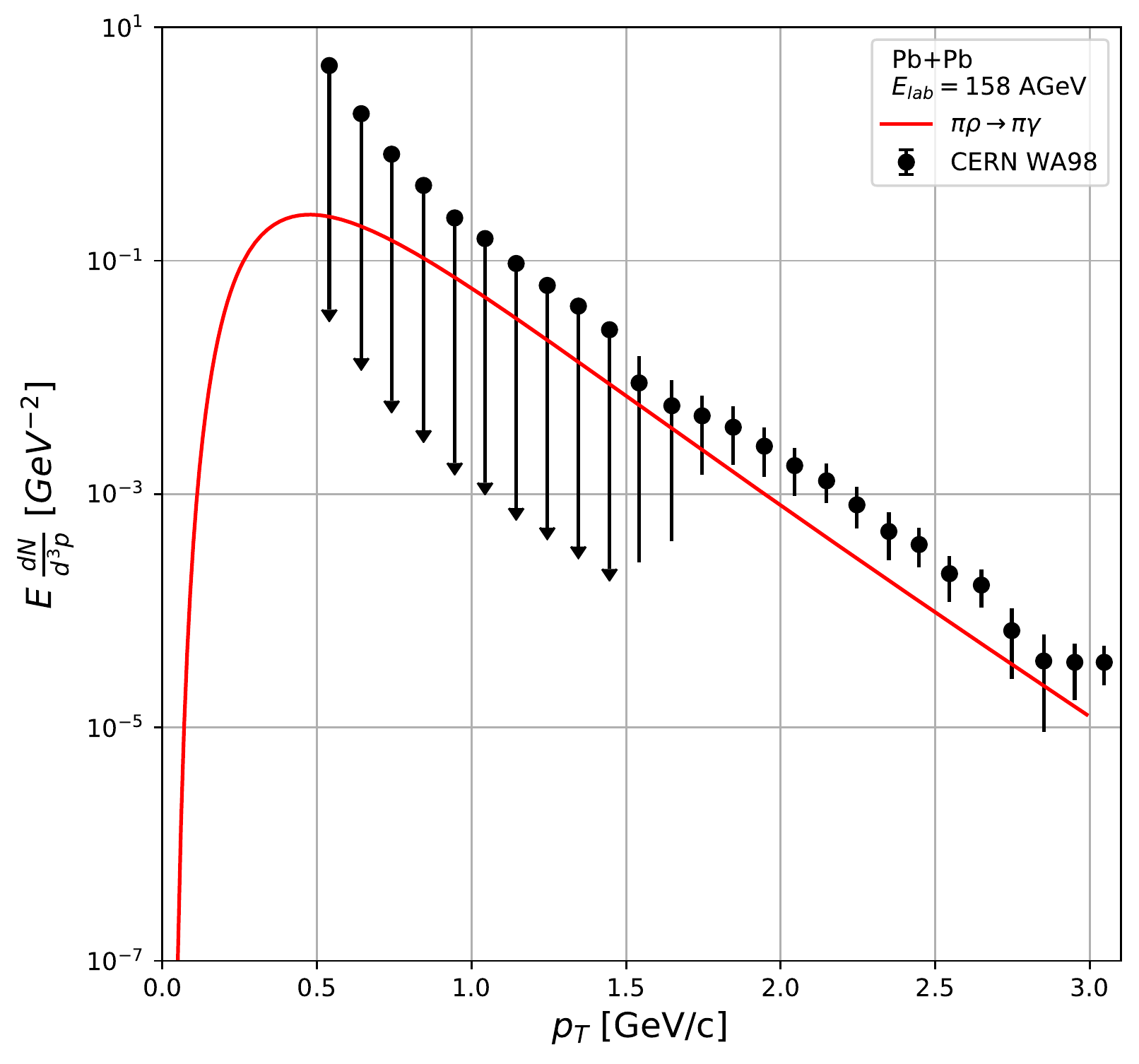}
\end{center}
\caption{[Color online] Invariant photon spectrum for a Pb+Pb collision with $E_{lab}=158 A$~GeV (red) and experimental data from the CERN WA98 experiment (black) \cite{WA98:2000vxl}. Error bars with downward arrows indicate unbounded upper limits.}
\label{fig:CERN_data} 
\end{figure}

In Fig.~\ref{fig:CERN_data} we show the photon production rate from the process $\pi\rho \rightarrow \pi\gamma$ from the simulation of a collision of two lead nuclei at a beam energy of $E_{lab}=158 A$~GeV in comparison to data from the CERN~WA98 experiment at the same beam energy. Our calculations show good agreement with the experimental data within the statistical and systematical errors of the measurement although we have only considered one of the numerous possible hadronic production channels for photons here. However, as was shown by TRG in Ref. \cite{Turbide:2003si} the reaction $\pi\rho \rightarrow \pi\gamma$ is the dominant source of thermal photons at large transverse momenta which explains why our rates are consistently shifted to slightly lower values compared to the data. At low transverse momenta, where one expects higher contributions of other production channels (see e.g. \cite{Turbide:2003si}), this effect is even more pronounced. 

%%%%%%%%%%%%%%%%%%%%%%%%%%%
%%%%%%%%%%%%%%%%%%%%%%%%

\section{Summary (and what can be improved)}
We have studied the production of photons from nuclear collisions at relativistic energies. We learned that thanks to their long mean free path length, photons are perfect candidates to probe the hot and dense regions of a fireball created in such a nuclear collision. Therefore, traces of a phase transition of the QCD matter to a QGP can possibly be observed in photon spectra.
We shortly discussed that there are numerous production channels of photons in such collisions and then focused on thermal photon production from the most important reaction ($\pi\rho \rightarrow \pi\gamma$).

As a tool we used (3+1) dimensional hydrodynamics which is a common method of modeling the evolution of QCD matter. Besides energy and momentum conservation as well as conservation of the baryon current we saw that we need an EoS (relation between energy density, pressure, and temperature) to close the system of equations. Here, we used a hadron resonance gas EoS where particles are modeled as point particles. For more details on aspects of the EoS that should be considered, see the improvements list below. We found that the photon yield is sensitive to the space-time volume of the matter created in such collisions and that it allows to determine the average temperature of the matter. After technical set-up, photon rates from Gale et al. were used to calculate the spectra of photons at different collision energies. 

We conclude that photons allow (as expected) to obtain a clear view on the fireball that is superior to the analysis of hadron spectra.

Possible improvements: It is clear that the time scale of this exploration did not allow to include all state-of-the-art physical phenomena and can only serve as a teaser for the interested reader. Possible improvements are:
\begin{itemize}
    \item 
    Information on further photon production rates are available, see e.g. \cite{Turbide:2003si} and can be included in the calculation.
    \item
    For a rigorous calculation, photon rates that depend also on the baryon density should be employed. A parametrization of the rates is available and can be taken from \cite{Heffernan:2014mla}. This is of crucial importance because low energy heavy ion collision can indeed create densities that are also found in the interior of neutron stars or in neutron star mergers (where one expects densities up to 3-6 times normal nuclear matter densities). This provides an interesting link between laboratory experiments to explore the EoS of QCD matter and the physics of neutron stars and gravitational waves (see e.g. \cite{LIGOScientific:2018cki} where first constraints on the QCD EoS from the observation of the gravitational wave GW170817 were presented and allowed to constrain the radii of the neutron stars and the pressure as a function of baryon density).
    \item
    The EoS can be refined to include e.g. excluded volume effects (repulsive interactions), a van-der-Waals gas or a phase transition to a deconfined (QGP) state. This involves the construction of such an EoS or a parametrization of the EoS from lattice QCD. Such a parametrization is available, see e.g. \cite{Motornenko:2020vqm}. Such modification will change the space time evolution of the created matter and may lead to modified photon spectra. Such modifications and how to learn about them from photon spectra are topics of current research. 
    \item 
    If a QGP EoS is used, one should also employ the exact photon rate for a quark gluon gas. This can be obtained from \cite{Kapusta:1991qp}.
    \item
    Last but not least, one needs to include the experimental cuts if one aims to compare to the data.
\end{itemize}

%%%%%%%%%%%%%%%%%%%%%%%%%%%%%%%%%%%%%%%%%%%%%%%%

\section{What can be learned from this project?}
The topic is centered around the physics of today's largest accelerator facilities and fits very well into an advanced nuclear physics lecture. It demonstrates how a well understood concept, namely thermal radiation (known from Planck's law), can be used to learn something about the interior of the matter produced in such collisions. This opens the possibility to engage the students in a general (or specific) discussion about the physics explored in large accelerators (e.g. the quark-gluon-plasma \cite{Veleiev:2003}, constituents of matter, creation of mass, relation to neutron stars, gravitational waves and element synthesis) but also raises their interest for more theoretical questions like the quark model of hadrons \cite{Hussar:1980} and gauge theories in general \cite{Kenyon:2008} that are at the theoretical basis of the physics explored with relativistic particle accelerators. A plethora of resources for Quantum-Chromo-Dynamics (QCD) can be found in \cite{Kronfeld:2010}. 

We have chosen a hydrodynamic approach \cite{Landau:1959} for this study, because most students have already an intuitive idea about fluids. In addition, such an approach allows to work with the known concepts of temperature and chemical potentials which relates this study to the statistical mechanics lectures and also to astrophysics questions like compact stars. Thus, the students can see the interconnection between knowledge obtained from lectures and its application in high energy physics and astrophysics \cite{Macher:2005,Sagert:2006} where concepts like the partition function are also used to obtain an Equation-of-State.

In more detail the students learn about the connection to the physics of Compton scattering \cite{Bartlett:1964,Wilkins:1992}, which they might have studied before in the case of electrons. They learn that Compton scattering is not only relevant in the field of electron-photon scattering, but is also present in the scattering of other spin-1 bosons with other particles (e.g. gluon+quark $\rightarrow$ photon+quark, where the gluon is converted to a photon or $\rho$+pion $\rightarrow$ photon+pion, where the $\rho$ meson was converted into a photon. Thus, the students learn that the Compton effect is present in a wider range of physics. In addition, one touched the question of massive (virtual) photons (e.g. in the case of the $\rho$ meson being converted to a photon) which allows for a general discussion of virtual particles. 

Apart from the physics skills, the students do acquire technical knowledge in solving a set of differential equations using a more complicated algorithm than is usually employed. 

\begin{acknowledgments}
HV wants to thank the group at Frankfurt for their support where this work was done as a summer study project. 
\end{acknowledgments}

%%%%%%%%%%%%%%%%%%%%%%%%%%%%%%%%%%%%%%%%%%%%%%%%%%%%%%%%%%%

\end{document}